\documentclass[preprint,12pt,authoryear]{elsarticle}

\usepackage{amssymb}
\usepackage{amsmath}
\usepackage{multirow}
\usepackage{xcolor}
\usepackage{tabu}

\journal{New Astronomy}

\def\astrobj#1{#1}

\begin{document}
\taburulecolor{black!15!white!85}

\begin{frontmatter}

\title{A New Analytic Galactic Luminosity Profile Function}

\author[uml]{D. Katz\fnref{fn1}}
\address[uml]{Department of Physics \& Applied Physics, University of Massachusetts Lowell, Lowell, MA 01852}

\author[ghs]{D. Kim}
\author[ghs]{M. Lenthall}
\author[ghs]{A. Merousis}
\author[ghs]{N. Sundaramurthy}
\author[ghs]{J. Kim}
\address[ghs]{Gardner High School, Gardner, MA 01440}

\fntext[fn1]{\begin{tabular}{l l}
P: 1-617-909-2006 \\
F: 1-978-934-3068 \\
E: daniel\_katz1@student.uml.edu
\end{tabular}}

\begin{abstract}
In 2010 Spergel introduced an alternative to the traditional Sersic form for galactic luminosity profiles based on modified Bessel functions of the second kind. His motivation was the desire for an accurate one-parameter profile form with a simple Fourier transform (in contrast to the Fourier transform of the Sersic profile which can't be written in closed form), but we have found that the Spergel profile almost universally makes integrals easier when it replaces the Sersic profile in the integrand. In the original paper on the subject Spergel noted that his profile seems to fit galaxies on average just as well as Sersic's. Here we make this observation quantitative by comparing the residuals from fitting Sersic and Spergel forms to data. We find that the Spergel profile actually fits \emph{better} than the Sersic for a random sample of $16$ galaxies.

\end{abstract}


\end{frontmatter}


\section{Introduction}
\cite{sersic} introduced a profile, which now bares his name, that has been the go-to function one uses for single-parameter fitting of galactic luminosity as a function of radius for many years. In addition to being a good empirical fit to many galaxy types its form,
\begin{equation}
I_{se} \propto e^{-x^{1/n}}
\label{eq:ser}
\end{equation}

\noindent where $n$ is parameter to be fitted and $x=R/R_0$ with $R$ the distance from the center of the galaxy and $R_0$ its half-light radius, is reasonably easy to interpret. A typical galaxy's observed luminosity profile, viewed on a log-log scale, appears to decay nearly linearly with distance. Thus we see that the parameter $n$ determines the steepness of this decay and its deviation from linearity. In eqn.~\ref{eq:ser} we use a proportionality sign instead of an equal sign to emphasize that there should be an overall factor which fixes the magnitude of the galaxy. Since we're only concerned with fitting the \emph{shape} of galactic luminosity profiles, we shall neglect this proportionality constant hereafter and just use equalities. Over the years there have been several generalizations of the Sersic form put forward which allow more flexability capable, \emph{e.g.} of fitting galaxies whose core structure is very distinct from their disk structure (\cite{truj}). However, this extra power comes at the expense of additional fitting parameters. Aside from applications where a functional form a galaxy's radial luminosity profile is needed, these fitting schemes are useful for the automated categorization of galaxies by partitioning the parameter space into galactic morphological types as in, \emph{e.g.}~\cite{morph}.\newline
\indent As outlined above the Sersic profile form is simple and accurate. On the other hand, it is highly unwieldly as part of an integrand. Consider, for instance, its Fourier transform, which comes up in the calculation of corrections due to seeing.
\begin{equation}
\mathcal{F}(I_{se}) = \int e^{-x^{1/n}}\sin(kR)dR
\end{equation}

\noindent Except for a few special values of $n$ this integral can't be written in closed form, though we recognize that there is no shortage of reliable numerical integration methods which can handle it. Another place where the Sersic profile sets up a road block in our analysis is in the deduction of dark matter halos around galaxies from their observed luminous matter distributions and rotation curves. This time we must integrate the luminous matter distribution (which we can reasonably assume is proportional to the luminosity profile) against Newton's gravitational potential. Assuming the galaxy is much wider than it is thick so that, in cylindrical coordinates, we can ignore the $z$-dependence of the matter distribution\footnote{This amounts to assuming that the matter distribution's $z$-dependence is Dirac's delta function $\delta(z)$.} which makes the potential energy of stars a distance $R$ from the center
\begin{equation}
V(R) = GM\int \frac{e^{-(R'/R_0)^{1/n}}}{|\vec{r}-\vec{r}'|}R'dRd'\phi'
\label{eq:pot}
\end{equation}

\noindent where $M$ is the galaxy's total (luminous) mass and $G$ is Newton's constant. Again, numerical techniques can certainly handle eqn.~\ref{eq:pot} but an analytic solution might lead to an exact relationship between $n$ and the parameters describing the dark matter distribution in, say, cusp/core models (several of which are described in~\cite{cuspcore}).

\section{Spergel's Profile}
\cite{spergel} has proposed the functional form
\begin{equation}
I_{sp} = \left( \frac{x}{2}\right)^\nu \frac{K_\nu (x)}{\Gamma(\nu +1)}
\label{eq:sperg}
\end{equation}

\noindent as a potential galactic luminosity profile. Here, $\Gamma$ is the Gamma function, $K_\nu$ is a modified Bessel function of the second kind\footnote{In his paper Spergel calls $K_\nu$ a ``modified spherical Bessel function of the third kind." This just corresponds to his use of a different naming convention for Bessel functions, and unfortunately this function is also variously  known as a ``modified Bessel function of the first kind."}, $\nu$ is the fitting parameter and we are suppressing an overall factor as we did with the Sersic profile. This form inherits its utility from the myriad identities and relations involving Bessel functions (\cite{ams55}). The Fourier transform of eqn.~\ref{eq:sperg} is
\begin{equation}
\int \left( \frac{x}{2}\right)^\nu \frac{K_\nu (x)}{\Gamma(\nu +1)}\sin(\vec{k}\cdot \vec{r})d\vec{r} = \frac{1}{2\pi}\frac{1}{(1+k^2R_0^2)^{\nu+1}}
\end{equation}

\noindent where $k^2 = k_x^2 +k_y^2$ for a two-dimensional transformation and $k^2 = k_x^2 +k_y^2 + k_z^2$ for a three-dimensional transformation. This follows by writing $\sin(kx) = kxJ_0(kx)$, with $J_0$ a Bessel function of the first kind, and using properties of Bessel funcitons. Similarly, the expansion in cylindrical coordinates
\begin{equation}
\frac{1}{|\vec{r}-\vec{r}'|}=\sum_{m=-\infty}^{\infty}\int_0^\infty dk J_m(kr)J_m(kr')e^{im(\phi-\phi')-k|z-z'|}
\end{equation}

\noindent allows us to evaluate exactly the potential energy of the luminous matter in a galaxy. Instead of eqn.~\ref{eq:pot} we have
\begin{eqnarray}
V(R) &=& \frac{GM}{2\pi R_0^2}\int \left( \frac{R'}{2R_0}\right)^\nu \frac{K_\nu (R'/R_0)}{\Gamma(\nu +1)}\frac{R'dR'd\phi '}{|\vec{r}-\vec{r}'|}\nonumber  \\
&=& \frac{GM}{2}\frac{\Gamma(-\nu -\tfrac{1}{2})}{\Gamma(\nu +\tfrac{3}{2})}\left(\frac{R}{2R_0}\right)^{2\nu+1} \label{eq:pot1} \\
&\times & _1F_2\left(\nu+1;\nu+\frac{3}{2},\nu+\frac{3}{2};\left(\frac{R}{2R_0}\right)^2\right) \nonumber \\
&+&\frac{GM}{2}\frac{\sqrt{\pi}\Gamma(\nu+\tfrac{1}{2})}{\Gamma(\nu+1)}\ _1F_2\left(\frac{1}{2};1,\frac{1}{2}-\nu;\left(\frac{R}{2R_0}\right)^2\right) \nonumber
\end{eqnarray}

\noindent in which $_1F_2$ are generalized hypergeometric functions. This formula is only valid when the value of $\nu$ is such that the gamma functions are non-singular. Admittedly, hypergeometric functions can sometimes be as difficult to evaluate numerically as the integral they're replacing, however they have many useful properties (\cite{ams55}) which aid the analysis of eqn.~\ref{eq:pot1}. Moreover, eqn.~\ref{eq:pot1} is of comparable complexity to the result of the same analysis with a ``zero"-parameter exponential model $\rho(\vec{r}) \propto \delta(z)e^{-R/R_0}$.\newline
\indent Included in (\cite{spergel}) are several figures qualitatively comparing the goodness-of-fit of the Sersic and Spergel profiles for a few sample galaxies. There we find that the best-fitting Sersic and the best-fitting Spergel are nearly the same curve, although the values of the parameters $n$ and $\nu$ don't coincide. In the next section we test the statistical significance of this observation.

\section{Data Selection, Preprocessing, and Fitting}
Our sample of galaxies was selected for a separate study of the galactic rotation curve problem and as such it consists entirely of galaxies for which rotation curve data is available. Because the members of this sample were picked without regard for the shape of their luminosity profiles it seems to us fair to use it here. All of our luminosity profile data comes from the Sloan Digital Sky Survey (SDSS) data release $10$ (\cite{sdss}). SDSS partitions extended objects into annuli of universal thickness (\emph{i.e.} the angular size of each annulus in does not depend properties of the object) and presents the average flux contained within each annulus over the $u,\ g,\ r,\  i,$ and $z$ bands. In this study we are only interested in fitting the shape of the galaxy's luminosity profile, so the units used to measure the flux and radii are immaterial. As SDSS recommends in its documentation, to ensure that all of our fitted profiles were conservative we radially integrated the mean flux, fit a spline to those numbers, and then differentiated the spline to obtain usable data points. Finally, these points were averaged over the bands to obtain the data we actually used for fitting Sersic and Spergel profiles. To aid the non-linear fitting algorithm both the radial and luminosity data were put through inverse hyperbolic sine transforms. Again because these data were originally retrieved by the authors for the purpose of a rotation curve study, each band was given a weight corresponding to the estimated mass-to-light ratio of the galaxy for that band where reliable estimates were available (\cite{gran}). Thus in some cases we are actually fitting functions to a galaxy's luminous mass profile rather than its luminosity profile. This hardly matters since the two are almost always assumed to be proportional and we are not interested in overall constants in our functions.\newline
\indent Table $1$ enumerates the galaxies used and shows the values of $\nu$ and $n$ which give the best fitting Spergel and Sersic profiles, respectively.

\begin{center}
\begin{table}
  \begin{tabu}{*{9}{|c}|}
\multicolumn{1}{c}{}\\[5pt]
    \multicolumn{1}{c}{}&\multicolumn{1}{c}{\astrobj{F563 V2}}&\multicolumn{1}{c}{\astrobj{UGC 5750}}&\multicolumn{1}{c}{\astrobj{NGC 4395}}&\multicolumn{1}{c}{\astrobj{F579 V1}}&\multicolumn{1}{c}{\astrobj{U6614}}&\multicolumn{1}{c}{\astrobj{UGC 4325}} \\
\hline
$\nu$ & $2.42$ & $1.88$ & $-0.08$ & $2.11$ & $2.38$ & $2.10$\\
\hline
$n$ & $1.07$ & $0.97$ & $0.89$ & $1.03$ & $1.20$ &  $1.06$ \\
\hline \multicolumn{1}{c}{}\\[5pt]
\multicolumn{1}{c}{}&\multicolumn{1}{c}{\astrobj{NGC 4062}}&\multicolumn{1}{c}{\astrobj{NGC 2742}}&\multicolumn{1}{c}{\astrobj{NGC 0701}}&\multicolumn{1}{c}{\astrobj{NGC 2608}}&\multicolumn{1}{c}{\astrobj{NGC 3495}}&\multicolumn{1}{c}{\astrobj{NGC 1087}} \\
\hline
$\nu$ & $3.84$ & $3.32$ & $3.54$ & $2.69$ & $3.09$ & $3.90$\\
\hline
$n$ & $1.35$ & $1.27$ & $1.30$ & $1.19$ & $1.20$ & $1.32$ \\
\hline 
  \end{tabu}
  \begin{center}
  \begin{tabu}{*{6}{|c}|}
\multicolumn{1}{c}{}\\[5pt]
\multicolumn{1}{c}{}&\multicolumn{1}{c}{}&\multicolumn{1}{c}{\astrobj{UGC 3672}}&\multicolumn{1}{c}{\astrobj{UGC 1421}}&\multicolumn{1}{c}{\astrobj{UGC 2715}}&\multicolumn{1}{c}{\astrobj{UGC 4321}}\\
\tabucline{2-6}
\multicolumn{1}{c|}{}&$\nu$ & $1.75$ & $4.00$ & $3.25$ & $3.37$ \\
\tabucline{2-6}
\multicolumn{1}{c|}{}&$n$ & $1.58$ & $1.33$ & $1.27$ & $1.25$\\
\tabucline{2-6}
\end{tabu}
\end{center}
  \caption{Best fitting parameters for the Spergel ($\nu$) and Sersic ($n$) profiles.}
\end{table}
\end{center}

\section{Results}
The simplest quantitative measure of goodness-of-fit is the sum of squared residuals between the data and the model (SSR). Since we search the parameter space with the goal of minimizing the SSR it is also a natural choice for comparing models. 
\begin{figure}[h!]
\begin{center}
\includegraphics[scale=0.7]{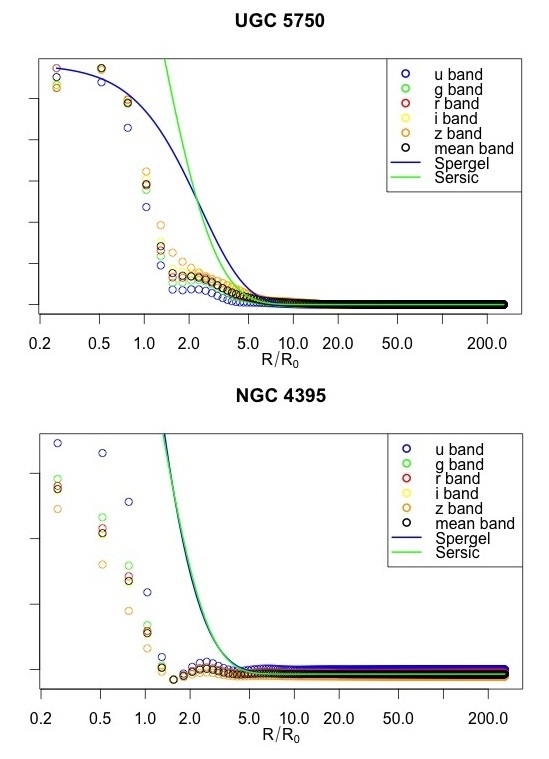}
\label{fig:gals}
\vspace{-0.5cm}
\caption{Raw band data in colors, averaged band data (to which the functions were actually fit) in black, and the fitting functions. (In the bottom panel the fitting functions lie almost directly on top of one another.)}
\end{center}
\end{figure}

\noindent After the preprocessing described in the previous section each galaxy was represented by the same number of data points. This means, combined with the fact that the Sersic and Spergel profiles have one degree of freedom each, that the SSR's don't need to be normalized in order to be comparable. Figure $1$ is a pair of galactic luminosity distributions, one of which is better fit by a Spergel profile while the other confounds both Sersic and Spergel functions. \astrobj{UGC 5750} is a good example of how Spergel's proflie can out perform Sersic's. A power times a Bessel function can be made to level off as $R \to 0$ while still having a modest slope as $R$ gets bigger. Also, this levelling off can be much more gradual than that of a decaying exponential allowing it to better mimic finite galactic cores. For the sake of fairness we also show \astrobj{NGC 4395}, which resists accurate fitting by both functions. This is probably because \astrobj{NGC 4395} is a dwarf galaxy of low surface bightness. In fact, it is the dimmest Seyfert galaxy yet found (\cite{ngc4395}). Cases where the Sersic profile fits better than Spergel were encountered, but we choose not to show their graphs because it is not clear by inspection that the SSR for one is lower than for the other. Figure $2$ shows histograms of the SSR's for each profile form.
\vspace{-0.5cm}
\begin{figure}[h!]
\begin{center}
\includegraphics[scale=0.7]{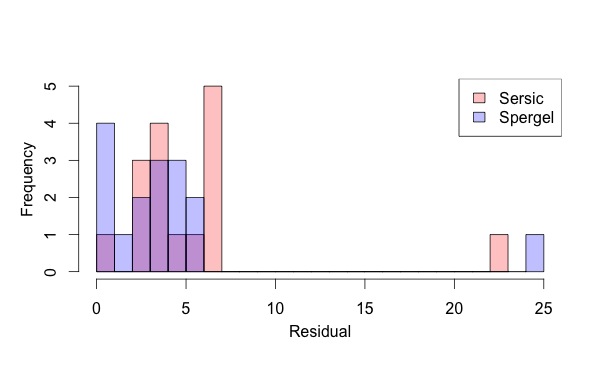}
\label{fig:hist}
\vspace{-1.5cm}
\caption{Distributions of SSR's over our sample set for the Spergel and Sersic profiles.}
\end{center}
\end{figure}

\noindent Here we see that the Spergel profile is not universally superior, but it does seem to favor lower SSR's. The two outliers correspond to the same galaxy (\astrobj{NGC 4395}) which neither profile fits especially well. Though Figure~\ref{fig:hist} is suggestive it is not visually obvious that it shows distinctly that the Spergel profile fits the sample generally better than the Sersic. For validation we perform a paired $t$-test on the SSR data (Spergel SSR's vs. Sersic SSR's) and find a $p$-value of about $0.013$.

\section{Conclusions}
In the course of using Spergel's Bessel-function-based luminosity profile for a galactic rotation curve study we have reinforced Spergel's claim that his profile works about as well as Sersic's on average. We quantified that sentiment using a paired $t$-test, based on which we can reasonably reject the hypothesis ``Spergel's profile fits on average worse than Sersic's." As a caveat we note that this rejection comes with all the usual complications of using a $p$-value for inference (see, for example,~\cite{pval}). Nevertheless, these results are evidence that the utility of Spergel's profile extends beyond its clean analytical properties.

\section*{Ackowledgements}
We are greatful to the SDSS for making their data publically and easily accessible. Additionally, we thank Dr. Zhaohui Yan of Gardner High School for providing us with space to work and for furnishing that space with computers. The lead author was supported during this work by the NSF's GK-12: Vibes \& Waves Fellowship [NSF \#0841392].
 


\section*{References}
\bibliographystyle{elsarticle-harv} 
\bibliography{spergbib}





\end{document}